\documentclass[9pt,conference]{IEEEtran}
\IEEEoverridecommandlockouts
\usepackage{booktabs}
\usepackage{amsmath,amssymb,amsfonts}
\usepackage{cite,url}
\usepackage{algorithmic}
\usepackage{graphicx}
\usepackage{textcomp}
\usepackage{xcolor}
\usepackage{bm}
\usepackage{paralist}
\usepackage{lineno}
\usepackage{pifont}
\usepackage{hyperref}

\def\BibTeX{{\rm B\kern-.05em{\sc i\kern-.025em b}\kern-.08em
    T\kern-.1667em\lower.7ex\hbox{E}\kern-.125emX}}
\begin{document}

\title{Music2Latent2: Audio Compression with Summary Embeddings and Autoregressive Decoding
\thanks{This work is supported by the EPSRC UKRI Centre for
Doctoral Training in Artificial Intelligence and Music (EP/S022694/1) and Sony Computer Science Laboratories Paris.}
}


\author{
\IEEEauthorblockN{Marco Pasini}
\IEEEauthorblockA{\textit{Queen Mary University} \\
London, UK}
\and
\IEEEauthorblockN{Stefan Lattner}
\IEEEauthorblockA{\textit{Sony Computer Science Laboratories} \\
Paris, France}
\and
\IEEEauthorblockN{Gy\"orgy Fazekas}
\IEEEauthorblockA{\textit{Queen Mary University} \\
London, UK}
}

\maketitle

\begin{abstract}
Efficiently compressing high-dimensional audio signals into a compact and informative latent space is crucial for various tasks, including generative modeling and music information retrieval (MIR). Existing audio autoencoders, however, often struggle to achieve high compression ratios while preserving audio fidelity and facilitating efficient downstream applications. We introduce Music2Latent2, a novel audio autoencoder that addresses these limitations by leveraging consistency models and a novel approach to representation learning based on unordered latent embeddings, which we call \textit{summary embeddings}. Unlike conventional methods that encode local audio features into ordered sequences, Music2Latent2 compresses audio signals into sets of summary embeddings, where each embedding can capture distinct global features of the input sample. This enables to achieve higher reconstruction quality at the same compression ratio. To handle arbitrary audio lengths, Music2Latent2 employs an autoregressive consistency model trained on two consecutive audio chunks with causal masking, ensuring coherent reconstruction across segment boundaries. Additionally, we propose a novel two-step decoding procedure that leverages the denoising capabilities of consistency models to further refine the generated audio at no additional cost. Our experiments demonstrate that Music2Latent2 outperforms existing continuous audio autoencoders regarding audio quality and performance on downstream tasks. Music2Latent2 paves the way for new possibilities in audio compression. 
\end{abstract}


\section{Introduction}
Representing high-dimensional audio data in a compact and informative latent space is valuable for various tasks, spanning generative modeling, music information retrieval (MIR), and audio compression.
While recent audio autoencoders have made significant strides in learning such representations, they still struggle to achieve high compression ratios while preserving audio fidelity and enabling downstream applications.
Existing approaches typically encode audio into ordered sequences of discrete tokens or continuous embeddings, where each element describes a short audio segment.
However, these methods inherently limit compression efficiency, as global audio features, such as timbre or tempo in the context of music samples, are redundantly encoded across multiple tokens or embeddings.
This work introduces Music2Latent2, a novel autoregressive audio autoencoder that overcomes these limitations by using unordered embeddings, which we call \textit{summary embeddings}: each summary embedding can capture distinct global features of a large chunk of the audio signal.
This is achieved by using learned embeddings and transformer blocks, and it allows for a more efficient allocation of information within the latent space, leading to higher reconstruction quality without compromising the compression ratio.
A consistency model that decodes the audio from latent embeddings is trained using causal masking in the self-attention layers, enabling it to attend to past audio segments during decoding, thus ensuring coherent reconstruction and avoiding boundary artifacts.
Furthermore, Music2Latent2 uses a novel two-step decoding procedure that exploits autoregressive decoding to achieve higher reconstruction quality without increasing computational cost.
Our experiments show that Music2Latent2 significantly outperforms continuous audio autoencoder baselines on audio quality of reconstructions at the same and at double the compression ratio, while achieving competitive results on MIR downstream tasks.

\begin{figure*}[t]
\centering
\includegraphics[width=0.85\textwidth]{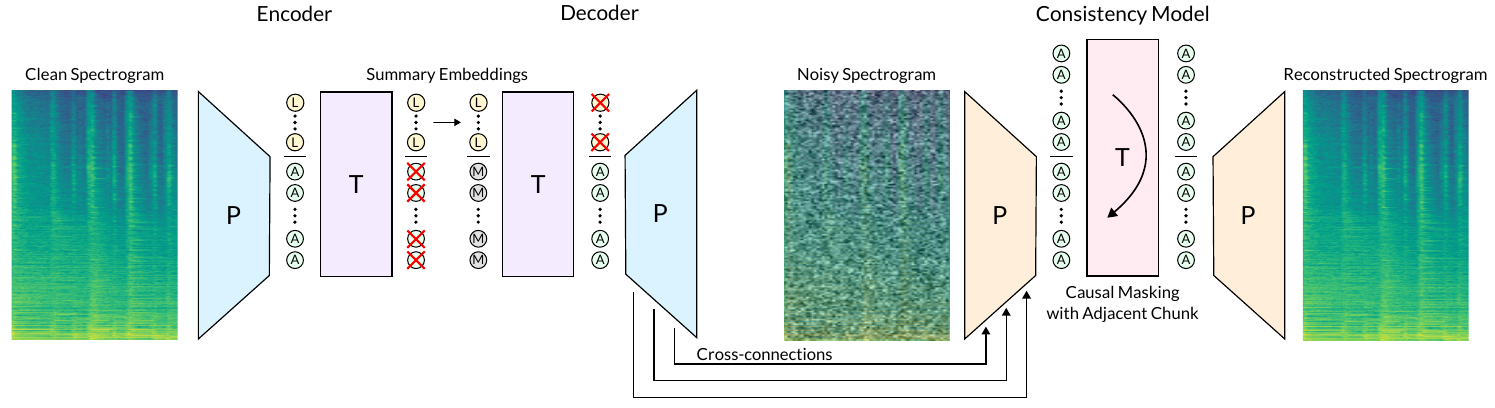}
\caption{Architecture of Music2Latent2. Convolutional patchifiers and de-patchifiers are indicated with \textit{P}, transformer modules with \textit{T}. Audio embeddings are illustrated as \textit{A}, learned/summary embeddings as \textit{L}, and mask embeddings as \textit{M}. We represent chunked causal masking with a curved arrow.}
\label{fig:architecture}
\end{figure*}

\section{Related Work}

\subsubsection{Audio Autoencoders}
The autoencoder used in Musika \cite{pasini_musika_2022} and the autoencoder proposed in \cite{bassnet} reconstruct the magnitude and phase components of a spectrogram, enabling fast inference but requiring a two-stage training process with an adversarial objective.
Stable Audio and Stable Audio 2 \cite{stable_audio, stable_audio_2, stable_audio_open} make use of audio autoencoders to produce latents for training generative models, but these autoencoders still rely on adversarial training and a careful balance between multiple loss terms.
Mo\^usai \cite{schneider_mousai_2023} introduces a diffusion autoencoder for learning an invertible audio representation, but while only using a single loss for training, inference requires multiple sampling steps.
Music2Latent \cite{music2latent} is a consistency autoencoder that is both trained with a single loss term and decodes samples in a single step, and it outputs an ordered sequence of latents.
SoundStream \cite{zeghidour_soundstream_2022}, EnCodec \cite{defossez_high_2022}, and Descript Audio Codec (DAC) \cite{kumar_high-fidelity_2023} encode samples to discrete codes using Residual Vector Quantization (RVQ). These models can achieve high fidelity reconstructions and are well-suited for training autoregressive models \cite{copet_simple_2023, jukebox, agostinelli_musiclm_2023}. 
%
They also operate at lower time compression ratios compared to the continuous counterparts, and are thus not directly comparable to our work.
The idea of using unordered embeddings to maximise the compression ratio has been successfully used in the vision domain for discrete autoencoders \cite{titok}.

\subsubsection{Consistency Models}
Consistency models \cite{song_consistency_2023, improvedconsistencysong} 
%
have shown impressive results in image generation tasks \cite{lcm}, achieving high-fidelity generation with single-step sampling.
The application of consistency models to audio generation remains relatively unexplored. CoMoSpeech \cite{comospeech} explores consistency distillation for speech synthesis, but relies on a pre-trained diffusion model.
Music2Latent \cite{music2latent} is the first autoencoder to successfully apply consistency models for audio compression and representation learning.

\section{Background}
Consistency models learn a mapping from any point on a diffusion trajectory to its origin, effectively reversing the diffusion process.
The ordinary differential equation (ODE) of the probability flow is introduced by \cite{song_denoising_2021} as: $\frac{dx}{d\sigma} = -\sigma\nabla_x \log p_\sigma(x)$ with $\sigma \in [\sigma_{\text{min}}, \sigma_{\text{max}}]$ and where
$p_\sigma(x)$ is the perturbed data distribution after adding Gaussian noise with standard deviation $\sigma$ to the original data distribution $p_{data}(x)$. $\nabla_x \log p_\sigma(x)$ is the score function \cite{song_estimating_gradients, song_score-based_2020, song_improved_score}. The ODE establishes a bijective mapping between a noisy sample $x_\sigma \sim p_\sigma(x)$ and $x_{\sigma_{\text{min}}} \sim p_{\sigma_{\text{min}}}(x) \approx x \sim p_\text{data}(x)$.
A consistency model $f_\theta(x_\sigma, \sigma)$ is trained to approximate the consistency function $f(x_\sigma, \sigma) \mapsto x_{\sigma_{\text{min}}}$, and is parameterised as: $f_\theta(x_\sigma, \sigma) = c_{\text{skip}}(\sigma)x_\sigma + c_{\text{out}}(\sigma)F_\theta(x_\sigma, \sigma)$,
where $F_\theta(x_\sigma, \sigma)$ is a neural network, and $c_{skip}(\sigma)$ and $c_{out}(\sigma)$ are differentiable functions that satisfy the boundary condition $f_\theta(x_{\sigma_{min}}, \sigma_{min}) = x_{\sigma_{min}}$.
%
Consistency training allows to train consistency models without a teacher diffusion model. It involves discretising the probability flow ODE using a sequence of noise levels $\sigma_{min} = \sigma_1 < \sigma_2 < ... < \sigma_N = \sigma_{max}$ and minimising the loss $\mathcal{L}_{\text{CT}} = \mathbb{E} \left[ \lambda(\sigma_i, \sigma_{i+1}) d\left(f_\theta(x_{\sigma_{i+1}}, \sigma_{i+1}), f_{\theta^-}(x_{\sigma_i}, \sigma_i)\right) \right]$,
where $d(x, y)$ is a distance metric, $\lambda(\sigma_i, \sigma_{i+1})$ is a loss scaling factor, $f_{\theta^-}$ is a stop-gradient version of $f_\theta$.
The consistency model $f_\theta(x, \sigma)$ generates a sample $x$ in one step from $z \sim \mathcal{N}(0, I)$ by computing $x = f_\theta(\sigma_\text{max} z, \sigma_\text{max})$. 

\begin{table*}[t]
\centering
\begin{minipage}[t]{0.3\textwidth}
\caption{Ablation study.}
\centering
\begin{tabular}{lcc}
\toprule
& $\text{FAD}_\text{clap}\downarrow$ & $\text{FAD}\downarrow$ \\
\midrule
w/o summary emb. & 0.0333 & 1.139 \\
w/ summary emb. & \textbf{0.0262} & \textbf{0.970} \\
\bottomrule
\end{tabular}
\label{tab:ablation}
\end{minipage} \hfill
\begin{minipage}[t]{0.69\textwidth}
\caption{Audio compression/quality metrics. Best and second-best are bolded and underlined.}
\centering
\begin{tabular}{lcccccc}
\toprule
& Compression Ratio & Stereo & SI-SDR $\uparrow$ & ViSQOL $\uparrow$ & $\text{FAD}_\text{clap}\downarrow$ & $\text{FAD}\downarrow$ \\
\midrule
\textit{DAC} & \textit{N/A} & \ding{55} & \textit{\textbf{9.48}} & \textit{\textbf{4.21}} & \textit{0.041} & \textit{0.966} \\
\midrule
Musika & 64x & \ding{55} & -25.81 & 3.80 & 0.103 & 2.308 \\
LatMusic & 64x & \ding{55} & -27.32 & 3.95 & 0.050 & 1.630 \\
Mo\^usai\_v2 & 64x & \ding{51} & -21.44 & 2.36 & 0.731 & 4.687 \\
Mo\^usai\_v3 & 32x & \ding{51} & -17.47 & 2.28 & 0.647 & 4.473 \\
Music2Latent & 64x & \ding{55} & -3.85 & 3.84 & 0.036 & 1.176 \\
StableAudio & 64x & \ding{51} & \underline{6.04} & \underline{4.08} & 0.107 & 1.017 \\
\midrule
Music2Latent2 & 64x & \ding{55} & -0.58 & 3.85 & \textbf{0.017} & \textbf{0.570} \\
$\text{Music2Latent2}_{\text{stereo}}$ & \textbf{128x} & \ding{51} & -2.29 & 3.91 & \underline{0.023} & \underline{0.717} \\
\bottomrule
\end{tabular}
\label{tab:audio_quality}
\end{minipage}
\end{table*}


\section{Music2Latent2}

\subsubsection{Audio Representation}
Similarly to \cite{music2latent}, Music2Latent2 uses complex-valued STFT spectrograms as the input representation for audio signals \cite{drumgan, comparing}. The 2D nature of spectrograms allows for the direct application of UNet \cite{unet} and DiT \cite{dit} architectures that have been successfully used in diffusion-based image generation.
We also use the spectrogram amplitude transformation from \cite{music2latent, speech_enhancement} to address the challenge of varying amplitude value distributions across frequencies. We treat the complex STFT spectrogram as a 2-channel representation, with each channel corresponding to the real and imaginary components, respectively.

\subsubsection{Architecture}
The architecture, as shown in Fig. \ref{fig:architecture}, is similar to Music2Latent and includes an encoder, a decoder, and a consistency model. Music2Latent has a convolutional architecture that allows the decoding of audio samples with different lengths than those used during training. In contrast, Music2Latent2 includes transformer blocks in the three modules, making the decoding of arbitrary-length audios challenging. This is due to the quadratic scaling of memory requirements of self-attention with increasing audio length and the difficulty of transformers generalising to sequence lengths different from those used during training. We thus propose to perform decoding via chunked autoregression, which allows us to use the same sequence length at both training and inference. All architecture components operate on independent chunks, except for the transformer blocks of the consistency model, which operate on two consecutive chunks with causal masking.

The \textbf{encoder} takes a spectrogram chunk as input and uses a convolutional patchifier to downsample it into lower time and frequency resolution patches. We then apply the technique proposed in TiTok \cite{titok}, appending a set of $K$ learnable latent embeddings to the flattened sequence of audio patches. This augmented sequence is then fed into a stack of transformer blocks, allowing the model to learn global relationships between audio features and the learnable latents.,
We then discard the audio embeddings, retaining only the $K$ resulting summary embeddings, which can now contain global information about the input chunk. A $tanh$ function is used to constrain the $K$ $d_{lat}$-dimensional embeddings in the $(-1,1)$ range \cite{pasini_musika_2022, schneider_mousai_2023, music2latent}.

The \textbf{decoder} mirrors the architecture of the encoder, and it takes as input a set of $K$ summary embeddings.
In place of the audio embeddings, learnable ``mask'' embeddings are concatenated.
This combined sequence is then processed by a stack of transformer blocks.
The resulting audio embeddings are kept and fed into a convolutional de-patchifier, which gradually upsamples them. The only goal of the decoder is to feed intermediate features at different resolutions to the patchifier of the consistency model via cross-connections.

The \textbf{consistency model} uses a patchifier, transformer blocks and a de-patchifier, in order to produce an output with the same shape as the noisy spectrogram given as input. There are additive skip-connections between each resolution level of the patchifier and de-patchifier. There are also additive cross-connections from the decoder to each level of the patchifier to ``leak'' to the model information from the summary embeddings.
As noted in \cite{music2latent}, we find this design choice crucial to decode in a single step. Since at inference the input to the consistency model is an uninformative fully noisy spectrogram, the model greatly benefits from access to semantic features about which sample to reconstruct at early layers of the architecture.
The transformer blocks accept audio embeddings from two consecutive chunks as input and perform chunked causal self-attention.

\subsubsection{Training Process}
We train Music2Latent2 on two consecutive spectrogram chunks $x$ of length \textit{spec\_length}. Each chunk is processed independently except for the transformer in the consistency model, where we concatenate the flattened sequence of both samples into a single sequence and use causal masking in the self-attention layers. This effectively teaches the model to condition the generation of the current audio segment on the preceding segment, resulting in a coherent reconstruction without boundary artifacts. We thus have:
\begin{align*}
    \hat{x}_{\text{left}},\hat{x}_{\text{right}} &= 
    \text{CM}_{\sigma_{\text{left}},\sigma_{\text{right}}}(\text{Dec}(\text{Enc}(x_{\text{left}})), x_{\text{left}}+\sigma_{\text{left}}\varepsilon_{\text{left}}, \\
    &\qquad \qquad \qquad \text{Dec}(\text{Enc}(x_{\text{right}})), x_{\text{right}}+\sigma_{\text{right}}\varepsilon_{\text{right}})
\end{align*}
where Enc, Dec, and CM are the Encoder, Decoder and Consistency Model, $\sigma \in [\sigma_{\text{min}}, \sigma_{\text{max}}]$ are noise levels and $\varepsilon \sim \mathcal{N}(0, I)$.
During training, we sample independent noise levels $\sigma_{\text{left}}$ and $\sigma_{\text{right}}$. This allows us to dynamically change the noise level of each chunk at inference.
We adopt the same EDM framework used by \cite{improvedconsistencysong} regarding the Pseudo-Huber loss $d(x,y)$ \cite{huber_loss} and the loss weighting $\frac{1}{\Delta\sigma}$:
\begin{equation*}
    \mathcal{L} = \mathbb{E} \left[ \frac{1}{\Delta\sigma} d\left(\text{CM}_{\sigma_{\text{left}}+\Delta\sigma,\sigma_{\text{right}}+\Delta\sigma}, \text{sg}\left(\text{CM}_{\sigma_{\text{left}},\sigma_{\text{right}}}\right)\right) \right]
\end{equation*}
where $\Delta\sigma$ is the step between adjacent noise levels and sg is the stop-gradient operator. We use this single loss to train the model end-to-end.
We also follow \cite{music2latent} and use continuous noise levels and an exponential consistency step schedule.

\begin{figure}[h]
\centering
\includegraphics[width=0.42\textwidth]{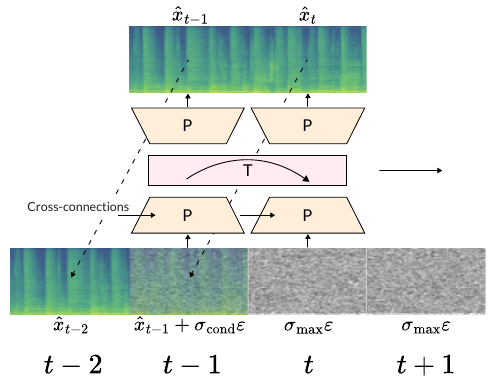}
\caption{Autoregressive decoding of Music2Latent2.}
\label{fig:inference}
\end{figure}

\subsubsection{Inference}
To \textbf{encode} an audio signal of arbitrary length, we first compute its spectrogram with temporal dimension $N\cdot \textit{spec\_len}$ (zero-padding if necessary). The spectrogram is then split into $T$ chunks. Each chunk is processed independently by the encoder, producing $T$ sets of $K$ summary embeddings.
Crucially, as the encoder operates on fixed-length chunks, the encoding process can be fully parallelised.

To \textbf{decode} we use an autoregressive approach, as shown in Fig. \ref{fig:inference}. First, the decoder produces cross-connections for each timestep $t$ from the corresponding $K$ summary embeddings.
The first chunk $\hat{x}_{t=0}$ is decoded independently in a single step by the consistency model, conditioned on the cross-connections.
For $\hat{x}_{t>0}$, the previously decoded $\hat{x}_{t-1}$ is corrupted with Gaussian noise at a controlled noise level $\sigma_{\text{cond}}$. Thereby, when decoding $\hat{x}_t$, $\hat{x}_{t-1}$ is decoded again, in the same model evaluation. With consistency models, generation quality often improves when sampling with more than a single step \cite{song_consistency_2023, improvedconsistencysong}, and we exploit this at no additional computational cost. By re-introducing noise in $\hat{x}_{t-1}$, we can also avoid the error accumulation characteristic of autoregressive models, especially if trained on continuous data \cite{rolling_diffusion}. The added noise introduces a degree of uncertainty into the past audio segment which results in the model not copying over the previously committed errors \cite{diffusion_forcing}.

\begin{table*}[t]
\caption{Downstream task results. Best results among autoencoder baselines are underlined.}
\centering
\begin{tabular}{lcccccccc}
\toprule
& \multicolumn{2}{c}{MagnaTagATune} & \multicolumn{2}{c}{Beatport} & \multicolumn{2}{c}{TinySOL-pitchclass} & \multicolumn{2}{c}{TinySOL-instrument} \\
& AUC-ROC & AUC-PR & Micro Acc. & Macro Acc. & Micro F1 & Macro F1 & Micro F1 & Macro F1 \\
\midrule
MusiCNN-MSD & 87.6 & 37.5 & 13.5 & 7.3 & 17.2 & 15.7 & 68.2 & 60.8 \\
CLMR & 89.9 & 42.6 & 13.9 & 7.8 & 16.8 & 16.2 & 93.5 & 89.7 \\
MERT-v1-95M & \textbf{90.8} & \textbf{44.9} & 50.7 & 44.3 & 98.3 & 98.3 & \textbf{97.1} & \textbf{95.8} \\
\midrule
Musika & 84.8 & 32.9 & 45.2 & 41.0 & 93.5 & 93.4 & 93.3 & 84.5 \\
LatMusic & 85.9 & 34.9 & 37.4 & 30.2 & 88.9 & 88.8 & 92.6 & 80.7 \\
Mo\^usai\_v2 & 86.2 & 35.4 & 48.2 & 42.0 & 95.1 & 95.1 & 82.8 & 68.6 \\
Mo\^usai\_v3 & 85.8 & 34.5 & 39.8 & 31.9 & 95.5 & 95.6 & 93.1 & 82.3 \\
Music2Latent & \underline{88.6} & 39.7 & 65.5 & 60.1 & \underline{\textbf{99.8}} & \underline{\textbf{99.8}} & 92.6 & 81.0 \\
StableAudio & 88.4 & 38.7 & 64.2 & 58.3 & \underline{\textbf{99.8}} & \underline{\textbf{99.8}} & 91.3 & 79.3 \\
\midrule
Music2Latent2 & \underline{88.6} & \underline{39.9} & \underline{\textbf{66.3}} & \underline{\textbf{61.5}} & \underline{\textbf{99.8}} & \underline{\textbf{99.8}} & \underline{96.4} & \underline{87.4} \\
\bottomrule
\end{tabular}
\label{tab:downstream}
\end{table*}

\subsubsection{Implementation Details}
The patchifiers and de-patchifiers are implemented using the same convolutional blocks as in Music2Latent \cite{song_score-based_2020}. We use sinusoidal embeddings with $256$ channels \cite{transformer} to represent the noise levels, taking $\frac{\log(\sigma)}{4}$ as input. We condition all consistency model layers on the noise level using AdaLN \cite{dit}.
All skip and cross-connections across the model are additive. For all patchifiers we use 5 resolution levels, adopting $[3,3,3,4,5,1]$ layers per level and $[64, 128, 256, 256, 256, 256]$ channels per level. The architecture of the de-patchifiers is mirrored. For each of the three modules we use $16$ pre-LN transformer blocks with $\textit{dim}=256, \textit{heads}=4, \textit{mlp\_mult}=4$. The resulting model has $\sim 100$ million parameters. The remaining hyperparameters regarding the EDM framework, Pseudo-Huber loss function, consistency step schedule and STFT spectrogram calculation/rescaling are the ones used in Music2Latent \cite{music2latent}.
We train the model on waveforms of $67,072$ samples, whose STFT spectrograms are then split in half along the time axis so each chunk has $\textit{spec\_length}=64$. We choose $d_{lat}=64$ and $K=8$, and the model thus produces summary embeddings of $44.1\, \text{kHz}$ audio at a sampling rate of $\sim 11\, \text{Hz}$, with a time and total compression ratio of $4096$x and $64$x, respectively. We train with $\textit{batch\_size}=16$ for 1M iterations using RAdam \cite{radam} ($\textit{lr}_0=1e^{-4}, \beta_1=0.9, \beta_2=0.999$). A cosine learning rate decay with $\textit{lr}_{\textit{final}}=1e^{-6}$ and an Exponential Moving Average (EMA) of the parameters with $\textit{momentum}=0.9999$ are used. Training takes $\sim 10$ days on a single A100 GPU.

\begin{figure}[h]
\centering
\includegraphics[width=0.42\textwidth]{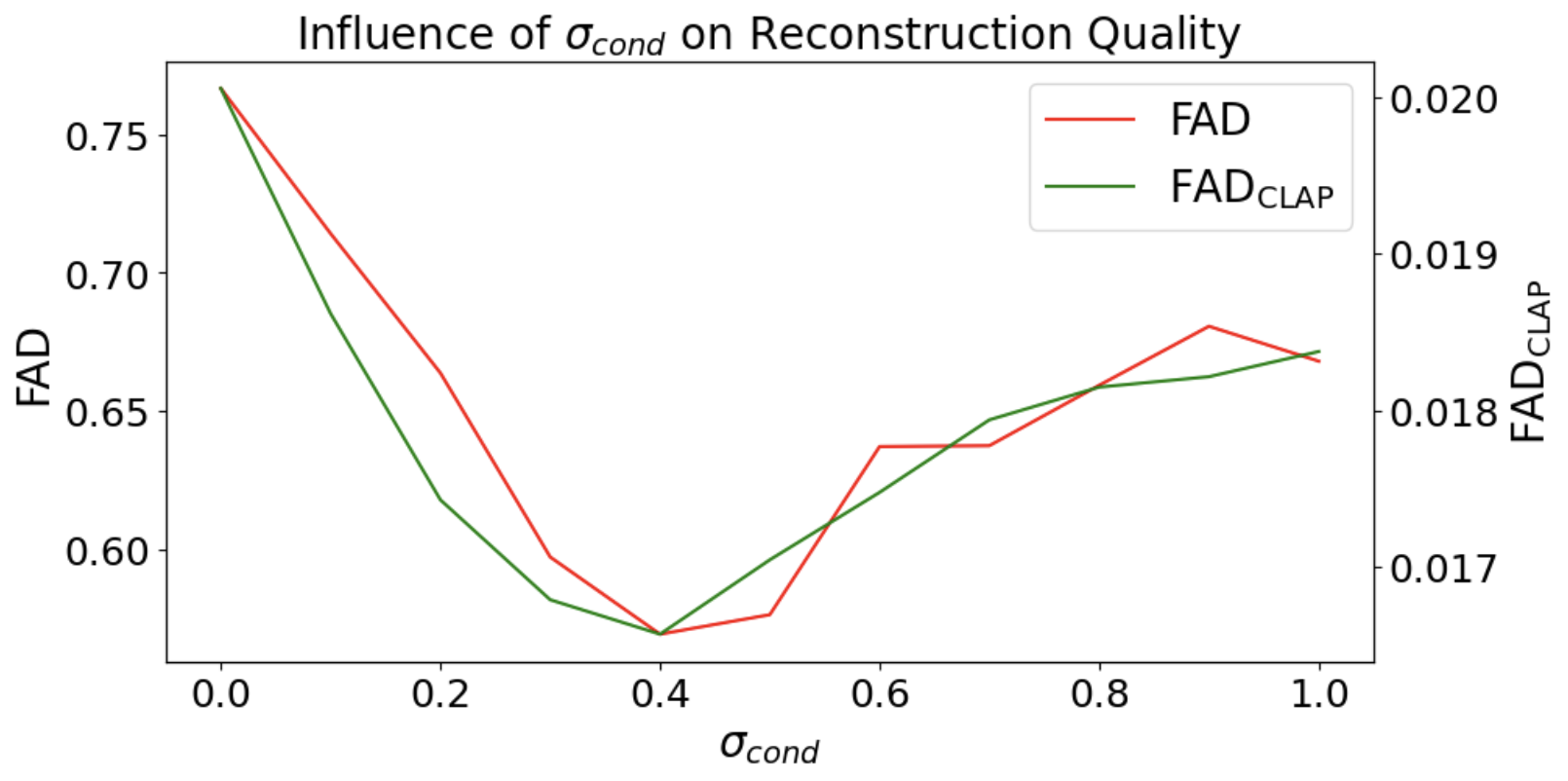}
\caption{Impact of $\sigma_{\text{cond}}$ on FAD and $\text{FAD}_\text{clap}$ for two-step decoding.}
\label{fig:noise_level}
\end{figure}
\vspace{-6pt}

\section{Experiments and Results}
\subsubsection{Experimental Setting}
We train Music2Latent2 on the same open datasets as Music2Latent: music from the MTG Jamendo dataset \cite{mtg_jamendo}, and speech from the DNS Challenge 4 dataset \cite{dns}, keeping the original sampling rates of $44.1\, \text{kHz}$ and $48\, \text{kHz}$ and sampling from each with equal weights.
We start from and expand on the evaluation framework originally proposed in \cite{music2latent} and we thus use MusicCaps \cite{agostinelli_musiclm_2023} as the evaluation dataset.
We choose the following continuous autoencoder baselines:
the autoencoder from \textit{Musika} \cite{pasini_musika_2022}, the autoencoder used in \cite{bassnet} to train an accompaniment generation model (we denominate it as \textit{LatMusic}), the \textit{v2} and \textit{v3} diffusion autoencoders used in \textit{Mo\^usai} \cite{schneider_mousai_2023}, the autoencoder used for \textit{Stable Audio Open} \cite{stable_audio_2, stable_audio_open} and \textit{Music2Latent} \cite{music2latent}, which uses a similar consistency framework to Music2Latent2 without relying on summary embeddings and transformer blocks.
We also include Descript Audio Codec (\textit{DAC}) \cite{kumar_high-fidelity_2023} for specific evaluations, even though not directly comparable. We provide audio samples at \href{https://anonymous2732.github.io/music2latent2/}{anonymous2732.github.io/music2latent2/}.

\subsubsection{Influence of $\sigma_{\text{cond}}$}
To investigate the influence of the noise level $\sigma_{\text{cond}}$ introduced during the two-step decoding procedure, we evaluate Music2Latent2 with different values of $\sigma_{\text{cond}}$ ranging from 0 to 1. Fig. \ref{fig:noise_level} shows the Frechét Audio Distance (FAD \cite{fad}) and $\text{FAD}_\text{clap}$ \cite{fad_correlation} (using CLAP \cite{clap} features) obtained for each noise level. The lowest FAD is achieved when $\sigma_{\text{cond}}=0.4$, and we thus use this value for all future experiments.


\subsubsection{Ablation Study}


To investigate the effectiveness of the summary embedding mechanism, we conduct an ablation study comparing Music2Latent2 against a variant where we do not concatenate any learned/summary embeddings in the transformers of the encoder/decoder, thus operating only on audio embeddings and producing a sequence of ordered latents.
To obtain the same number of compressed embeddings with the same dimensionality (which results in the same compression ratio), we apply a linear layer to the output of the transformer for each timestep. The two variants differ only with respect to this aspect, and are trained for 600k iterations using $[2,2,2,2,2,1]$ layers per level. The remaining architecture and training parameters are unchanged. We report FAD in Tab. \ref{tab:ablation}, where we show that using summary embeddings results in lower FAD.

\subsubsection{Audio Compression and Quality}
We use the evaluation framework as in \cite{defossez_high_2022, music2latent}, which consists of SI-SDR \cite{sisdr}, ViSQOL \cite{visqol, visqolaudio, visqolv3}, FAD and $\text{FAD}_\text{CLAP}$. SI-SDR and ViSQOL directly compare reconstructions to the original samples, while FAD-based metrics evaluate the audio quality of reconstructions without relying on pairs. In the case of an ideal autoencoder, increasing the compression ratio would result in a decrease of pair-wise metrics, since less information travels through the bottleneck. On the other hand, audio quality metrics would remain constant, since the missing information would be realistically generated while decoding. In the comparison we also include a $\text{Music2Latent2}_{\text{stereo}}$ model trained on stereo samples (the input is composed of the spectrograms of the two channels concatenated channel-wise), keeping the remaining architecture and training parameters unchanged.
In Tab. \ref{tab:audio_quality} we show that Music2Latent2 substantially outperforms all baselines in terms of FAD. DAC and the Stable Audio Open autoencoder perform better in terms of pair-wise metrics. A likely explanation is that these models are trained using several reconstruction losses (differences between output and input are directly penalised), while Music2Latent2 is trained purely as a generative model using a single consistency loss function, which does not directly compare reconstructions to the inputs. $\text{Music2Latent2}_{\text{stereo}}$ surpasses in FAD baselines with half of its compression ratio. We use single-step decoding for Music2Latent, since the audio quality deteriorates when using more than a single step \cite{music2latent}.

\subsubsection{Downstream Task Performance}
We investigate the effectiveness of Music2Latent2's latent representations for downstream MIR tasks by conducting experiments on three standard benchmarks: \textit{MagnaTagATune} \cite{magnatagatune} for autotagging, \textit{Beatport} \cite{beatport} for key estimation, and \textit{TinySOL} \cite{tinysol} for pitch and instrument classification. Embeddings are extracted before the last linear layer of the encoder for most models, where the number of channels is the highest, and then averaged across the time dimension. For Music2Latent2, we gather the summary embeddings for each chunk before the last linear layer of the transformer section in the encoder and stack them along the channel dimension. We then average the resulting embedding across the different chunks of the input. We also include in the evaluation common representation learning baselines: MusiCNN-MSD \cite{musicnn}, CLMR \cite{clmr}, and MERT-v1-95M \cite{repr_mert}. We adopt the same testing methodology adopted by \cite{music2latent, christos_thesis} using the \texttt{mir\_ref} library \cite{mir_ref}.
Tab. \ref{tab:downstream} shows how Music2Latent2 beats all autoencoder baselines across all metrics and is even superior to state-of-the-art representation learning models for key and pitch-class estimation. These results can be motivated by the use of summary embeddings in Music2Latent2, which can encode global features about the input sample and can thus result in a higher degree of feature disentanglement. We plan to explore this further in future work.

\section{Conclusion}
This work introduced Music2Latent2, a novel autoregressive audio autoencoder leveraging summary embeddings and consistency models for high-fidelity audio compression. By encoding audio into sets of summary embeddings, Music2Latent2 achieves higher reconstruction quality at the same compression ratio compared to conventional ordered embedding approaches. The autoregressive design enables processing of arbitrary-length audio signals while maintaining coherence and avoiding boundary artifacts. A two-step decoding process further improves the quality of reconstructions at no additional cost.
Our experiments demonstrate Music2Latent2's superior performance over existing continuous audio autoencoders on both reconstruction audio quality metrics and downstream MIR tasks. Music2Latent2 opens novel possibilities for neural audio compression and efficient generative modeling.

\bibliographystyle{IEEEtran}
\bibliography{refs}

\end{document}